\documentclass[paper]{JHEP3} 

\JHEPspecialurl{http://jhep.sissa.it/JOURNAL/JHEP3.tar.gz}

\usepackage{epsfig,multicol,bbm}

\newcommand\fverb{\setbox\pippobox=\hbox\bgroup\verb}
\newcommand\fverbdo{\egroup\medskip\noindent%
			\fbox{\unhbox\pippobox}\ }
\newcommand\fverbit{\egroup\item[\fbox{\unhbox\pippobox}]}
\newbox\pippobox

\title{Holographic Quantum Statistics \\ from Dual Thermodynamics}

\author{Csaba Bal\'azs\\ 
             HEP Division, Argonne National Laboratory,\\
             9700 Cass Ave., Argonne, IL 60439, USA\\
  	    E-mail: \email{balazs@hep.anl.gov}}
\author{Istv\'an Szapudi\\
             Institute for Astronomy, University of Hawaii, \\
             2680 Woodlawn Dr., Honolulu, HI 96822, USA\\
            E-mail: \email{szapudi@ifa.hawaii.edu}}

\preprint{ANL-HEP-PR-06-43}

\abstract{

We propose dual thermodynamics corresponding to black hole mechanics with 
the identifications $E' \to A/4$, $S' \to M$, and $T' \to T^{-1}$ in Planck 
units. Here $A$, $M$ and $T$ are the horizon area, mass and Hawking 
temperature of a black hole and $E'$, $S'$ and $T'$ are the energy, entropy 
and temperature of a corresponding dual quantum system. We show that, for a 
Schwarzschild black hole, the dual variables formally satisfy all three laws 
of thermodynamics, including the Planck-Nernst form of the third law 
requiring that the entropy tend to zero at low temperature. This is in 
contrast with traditional black hole thermodynamics, where the entropy is 
singular. Once the third law is satisfied, it is straightforward to 
construct simple (dual) quantum systems representing black hole mechanics. 
As an example, we construct toy models from one dimensional (Fermi or Bose) 
quantum gases with $N \simeq M$ in a Planck scale box. In addition to 
recovering black hole mechanics, we obtain quantum corrections to the 
entropy, including the logarithmic correction obtained by previous papers. 
The energy-entropy duality transforms a strongly interacting gravitational 
system (black hole) into a weakly interacting quantum system (quantum gas) 
and thus provides a natural framework for the quantum statistics underlying 
the holographic conjecture.

}


\begin{document} 


\section{Introduction}

The quantum estimate of the cosmological vacuum energy density is about 123 
orders of magnitude larger than its measured value 
\cite{Weinberg:1988cp,Riess:1998cb}.  This discrepancy might be resolved by 
an upper limit on the entropy of the universe, since such a bound implies an 
upper limit on its total energy density.  Along the lines of this argument, 
recently we conjectured that after imposing a holographic entropy limit, the 
resulting quantum theory leads exactly to the measured value of the total 
energy density \cite{Balazs:2006kc}.  In this work we construct an example, 
where imposing the holographic entropy limit on a quantum system indeed 
leads to an upper limit on its energy density. We choose a black hole as a 
representative holographic quantum system, since it is easy to relate the 
entropy and energy using black hole mechanics.

Black hole mechanics is one of the most tantalizing features of general 
relativity. Its formal analogy with thermodynamics was noticed early on 
\cite{Bekenstein:1973ur,Bekenstein:1974ax,Bardeen:1973gs}. The quantum 
nature of the underlying physics was clarified by the discovery of the 
Hawking radiation \cite{Hawking:1974rv,Hawking:1974sw}. While the quantum 
theory of black holes struggles with unsolved problems such as the 
information paradox \cite{Susskind:1993if}, it successfully motivated the 
holographic conjecture 
\cite{Bekenstein:1972tm,'tHooft:1993gx,Susskind:1994vu}: the entropy in a 
spherical volume cannot exceed the quarter of its surface area in Planck 
units (see \cite{Bousso:1999xy,Bousso:2002ju} for alternative and 
generalized definitions). Despite the general belief that the explanation of 
the holographic conjecture must be of quantum nature, it is non-trivial to 
find simple quantum systems representing black hole mechanics. In this 
paper, we set out to construct a quantum statistical model displaying the 
holographic behavior of a Schwarzschild black hole.

Following \cite{Wald:1999vt}, we summarize the laws of black hole mechanics 
for an isolated Schwarzschild black hole (i.e. one without electric 
charge or angular momentum) characterized with its mass $M$, surface gravity 
$\kappa = 1/(4M)$, and surface area $A$.
\begin{itemize}
 \item {\em Zeroth Law:} $\kappa$ is the same everywhere on the 
horizon in a time independent black hole.
 \item {\em First Law:} 
  \begin{equation}
    \delta M = \frac{\kappa}{8\pi}\delta A
  \label{Eq:BHD1}
  \end{equation}
 \item {\em Second Law:} 
  \begin{equation}
    \delta A \ge 0
  \end{equation}
\end{itemize}

Based on the analogy with thermodynamics one associates the temperature 
with the surface gravity $T \to \kappa/2\pi$, the energy with the mass $E 
\to M$, and the entropy with the horizon area $S \to A/4$.  Holography in 
this context amounts to the conjecture that a black hole maximizes the 
entropy \cite{'tHooft:1993gx}.
The traditional assignment of variables is further motivated by the Hawking 
effect: a black hole emits black body radiation with temperature 
$\kappa/2\pi$. A black hole weakly interacting with its surrounding appears 
to obey the generalized version of the second law where the sum of the 
entropies of matter and the black hole always increases 
\cite{Bekenstein:1974ax}.

On the other hand, black hole evaporation due to Hawking radiation raises 
questions: is the unitary nature of quantum mechanics broken? It 
appears that an initial pure quantum state will evolve into a mixed thermal 
state through the process of black hole formation and evaporation; this 
amounts to information loss. In addition, it is not clear precisely what 
quantum degrees of freedom are responsible for the entropy of the black 
hole.

The resolution of the information paradox is still open.  Either quantum 
gravity allows correlations to be restored through the horizon, in contrast 
with classical gravity, thus information could be restored during 
evaporation; or quantum mechanics needs to be modified to allow non-unitary 
evolution \cite{Banks:1983by,Wald:1995yp,Hartle:1996rp}. The degrees of 
freedom responsible for the entropy have stimulated vigorous research. 
Competing models involve weakly coupled string states 
\cite{Strominger:1996sh,Peet:1997es}, conformal symmetry 
\cite{Aharony:1999ti,Carlip:2000nv,Carlip:2006fm}, 
string fuzzballs \cite{Mathur:2005zp} 
spin network states at \cite{Ashtekar:1997yu} or inside \cite{Livine:2005mw} 
the horizon, ``heavy'' degrees of freedom in induced gravity 
\cite{Frolov:1997up}, or non-local topological properties of black hole 
spacetime \cite{Hawking:1998jf}. In each of these models the entropy is 
proportional to the horizon area, yet none of them offer a clear and 
comprehensive picture associated with black hole thermodynamics.

Another peculiarity of black hole mechanics is rarely discussed: it violates 
the (strong version of the) third law of thermodynamics \cite{Wald:1997qp}. 
The entropy is in fact singular (tends to infinity rather than zero) as the 
temperature approaches zero. Since virtually every known quantum mechanical 
system in nature obeys the third law, we argue that this singularity 
expresses  yet another way the difficulty of quantizing gravity.

In order to construct quantum models representing black hole mechanics, we 
propose dual thermodynamics with the following assignments $E' \to A/4, S' 
\to M$, and $T' \to 2\pi/\kappa$ in Planck units. In the dual description, 
the entropy and energy exchange roles, and the inverse of the Hawking 
temperature plays the role of the temperature. By definition, the dual 
variables obey the first law while the validity of the second law follows 
from the fact that the mass of a Schwarzschild black hole can only increase 
if we neglect Hawking radiation. Note that at this point we do not aim to 
model the generalized second law which includes the sum of matter and black 
hole entropies. The advantage of our proposal is that  the dual entropy of a 
black hole is proportional to its dual temperature $S' \propto T'$, i.e. the 
third law of dual thermodynamics is satisfied as well. 

In the next section we formalize the duality transformation between two 
thermodynamical systems, and in Section \ref{Sec:DualStat} we construct a 
simple quantum model which reproduces the dual thermodynamics. In the last 
section we summarize our results after transforming back from the dual 
variables and compare the quantum corrections to the entropy with those 
available in the literature.

\section{Energy-entropy duality}
\label{Sec:TheDua}

We define the energy-entropy duality by assuming that the first
law in entropy representation corresponds to the first law in
dual energy representation.
We allow for pressure and chemical potential for full generality, although 
we exclude for the moment other entities, such as electric charge, 
magnetic susceptibility, etc. The first law in both representations reads
\begin{eqnarray}
  dS  &=& \frac{1}{T}dE+\frac{p}{T}dV-\frac{\mu}{T}dN, \\
  dE' &=& T'dS' - p'dV' + \mu' dN',
\end{eqnarray}
where $S, E, V, N, T, p, \mu$ are the thermodynamic entropy, energy,
volume, species number, temperature, pressure, and chemical potential,
respectively, and primed symbols denote dual thermodynamical variables.
Comparison of the two equations motivates the definition of the
entropy-energy duality
\begin{eqnarray}
  S \to E', ~~~~~
  E \to S', ~~~~~
  T \to \frac{1}{T'}, ~~~~~
  \frac{\mu}{T} \to -\mu'.
  \label{eq:eduality}
\end{eqnarray}
We assumed that the degrees of freedom $N$ is the same in both spaces, which 
fixes the transformation of $\mu$. From these transformation rules it
follows that  $\frac{p dV}{T} \to -p' dV'$ from the first law.

While we presented a definition which might be more generally applicable,
we aim specifically at application to black hole mechanics. As shown at the
end of the previous section, for vanishing chemical potential, and
$p dV = 0$, all three dual thermodynamic laws will be satisfied as
a consequence of black hole mechanics.
For the more general case with non vanishing chemical potential, 
we can show that it is sufficient to have $p' dV'=0$, $\mu > 0$ and $N \sim 
M$ for the dual second law to hold.  In this case 
 \begin{equation}
   dS' = \frac{1}{T'}(dE' + p'dV' - \mu'dN') \ge 0,
 \end{equation}
since $dE' = dS \ge 0$ and $- \mu'dN' \ge 0$ separately.  We will show that
the specific quantum statistics we propose to represent the dual
thermodynamics satisfies these sufficient conditions. 

The above arguments show that it is reasonable to assume that 
the proposed entropy-energy duality is meaningful for the specific
case of black hole mechanics, as all three laws of thermodynamics
hold in the dual variables. The precise necessary conditions for the
applicability of the entropy-energy duality in a more general case are
less clear, and are beyond the scope of the present paper.

Our duality proposal was motivated by the third law of thermodynamics.
The behaviour $S' \propto T'$ and $E' \propto T'^2$  in dual variables
is now easily reproducible by ordinary quantum statistics.  In fact, probably
multiple quantum systems could satisfy these sensible relations.  We 
show in the next section, that one dimensional quantum gases (whether Fermi 
or Bose) provide a fully consistent dual quantum statistical model for 
the dual black  hole thermodynamics.

\section{Dual quantum statistics: one dimensional quantum gas}
\label{Sec:DualStat}

In this section, we summarize the main results for the statistical mechanics 
of one dimensional quantum gases.  As we show in the next section, they 
provide a representation of dual black hole thermodynamics in the limit 
where $|\mu'| \ll T'$, with $\mu'$ corresponding to quantum corrections. All 
equation in this section are written in dual variables, therefore we omit 
the primes for convenience.

The energy, pressure and particle number of one dimensional quantum gases 
are calculated as
\begin{eqnarray}
E = pL &=&\frac{g L}{2\pi }\int_0^\infty f(\epsilon) \epsilon \; d\epsilon, 
\label{Eq:EpL} \\
N      &=& \frac{g L}{2\pi }\int_0^\infty f(\epsilon) \; d\epsilon,
\end{eqnarray}
where $f(\epsilon) = 1/(e^{(\epsilon-\mu)/T}\pm 1)$.  Hereafter the upper 
(lower) sign represents the Fermi-Dirac (Bose-Einstein) distribution.  The 
coefficient $g$ gives the internal degrees of freedom of the gas particles, 
and $L$ is the size of the system.  
The entropy is obtained from the integral form of the first thermodynamic 
law
\begin{equation}
   S = \frac{1}{T}(E+p L - \mu N).
\end{equation}

The above integrals can be expressed in terms of the polylogarithm function
$\rm{Li}_n(z) \equiv \sum_{k=1}^{\infty} z^k/k^n$:
\begin{eqnarray}
  E = p L &=& \mp \frac{g L}{2 \pi} \; {\rm Li}_2\left(\mp e^{\frac{\mu}{T}}\right) T^2, \\
  N       &=& \pm \frac{g L}{2 \pi} \log \left(1 \pm e^{\frac{\mu }{T}}\right) T.
\end{eqnarray}
We will use the expansions of these formulae in the limit of $|\mu| \ll T$. 
For a Fermi gas
\begin{eqnarray}
  E = p L &=& \frac{1}{24} g L \pi  T^2
   +\frac{g L}{2 \pi} \log (2) T \mu 
   +\frac{g L}{8 \pi } \mu ^2
   +O\left(\mu ^3\right), \label{Eq:EpF} \\
  N &=& \frac{g L}{2 \pi } \log (2) T
   +\frac{g L}{4 \pi } \mu 
   +\frac{g L}{16 \pi T} \mu ^2
   +O\left(\mu ^3\right), \\
  S &=& \frac{1}{12} g L \pi {T}
   +\frac{g L}{2  \pi } \log(2) {{\mu }} 
   + O(\mu _{D}^{3}),
  \label{Eq:SpF}
\end{eqnarray}
and for a Bose gas
\begin{eqnarray}
   E = p L &=&\frac{1}{12} g L \pi  T^2
   - \frac{ g L}{2 \pi } \left(\log \left(-\mu/T\right)-1\right) T \mu 
   - \frac{g L}{8\pi } \mu ^2 
   + O\left(\mu ^3\right), 
   \label{Eq:EpB} \\
   N &=&-\frac{g L}{2 \pi } \log \left(-\mu/T\right) T
   - \frac{g L}{4 \pi } \mu 
   - \frac{g L}{48} \frac{\mu ^2}{\pi T}
   + O\left(\mu^3\right), \\
   S &=&\frac{1}{6} g L \pi T 
   - \frac{g L}{2 \pi } (\log (-\mu/T ) - 1) \mu
   + O\left(\mu ^3\right).
   \label{Eq:SpB}
\end{eqnarray}

\section{Quantum statistical model of a black hole}

In this section we use the previous results to show that
dual black hole thermodynamics is represented by one dimensional quantum
gases.  Note that we return to primed notation for dual variables. 
In addition, we derive quantum 
corrections to the black hole mechanics based on our toy model.
When the  dual temperature, which is proportional to the mass 
of the black hole, 
approaches the Planck scale, we expect quantum corrections. In the dual 
space quantum effects will manifest themselves when $T'$ approaches 
$\mu'$, therefore we can assume that $\mu'$ is at most of the 
order of the Planck mass or smaller.  In that case, 
for a large black hole, $\mu N$ will be completely negligible.

We can set $g L$ of our dual quantum theory to ensure consistency 
of the energy with that of the black hole:
\begin{equation}
  g L = \frac{3}{2 \pi^2}, \, \, \frac{3}{4 \pi^2},
  \label{Eq:gL}
\end{equation}
for Fermi and Bose systems, respectively.  This means that the one 
dimensional dual quantum system lives in a space of size order of Planck 
length.

Equation (\ref{Eq:gL}) fixes the parametric freedom in our quantum models. 
Transforming back from the dual variables, the leading terms of Eqs. 
(\ref{Eq:SpF}) and (\ref{Eq:SpB}) reproduce the black hole energy (assuming 
$\mu$ is small).  Meanwhile, the leading terms of Eqs. (\ref{Eq:EpF}) and 
(\ref{Eq:EpB}) reproduce the black hole entropy as demonstrated by the 
leading terms of Eqs. (\ref{Eq:SFexp}) and (\ref{Eq:SBexp}).  This itself is 
a non-trivial result.  After fixing $g L$ to obtain the correct coefficient 
of the black hole energy, the dual system does not have any freely 
adjustable parameters.  Yet, not only the temperature (or mass) dependences 
of the black hole energy and entropy are reproduced, but also the 
coefficient of the entropy term is given correctly. Finally, from $dL = 0$ 
follows $p dV = 0$, which means that black hole mechanics with its holographic
aspects is perfectly represented by our dual model.

For the Fermi model, we also obtain that the number of degrees of freedom
\begin{equation}
  N = \frac{6 \log(2)}{\pi^2} M,
\end{equation}
is essentially given by the black hole mass in Planck mass units.
This is an entirely sensible prediction: the number of degrees 
of freedom grow extensively with the black hole mass, 
while our model faithfully reproduces the 
holographic growth of entropy. 

For the Bose model, $N$ has
a logarithmic sensitivity for $\mu'$, but for a wide range of
values $N \simeq M\log M$ will hold, including the
value we suggest from consistency with previous quantum
gravity calculations for the entropy correction. Note that we
do not discuss the possible effects Bose condensation on $N$,
as we assumed that $\mu' \ll T'$.

Next, we calculate quantum corrections to the black hole entropy considering 
corrections to the energy of the dual quantum gas. The entropy according to 
our Fermi model is
\begin{eqnarray}
  S &=& -\frac{48}{\pi}{\rm Li}_2(-e^{-\mu}) M^2 = \\
  & & 4 \pi M^2 
  - \frac{48 }{\pi} \log(2) M^2 \mu 
  + \frac{12 }{\pi} M^2 \mu^2  
  - \frac{ 2 }{\pi} M^2 \mu^3 
  + O(\mu^4),
  \label{Eq:SFexp}
\end{eqnarray}
while the Bose model gives
\begin{eqnarray}
  S &=& \frac{24}{\pi} \, {\rm Li}_2(e^{-\mu }) \, M^2 = \\
    & & 4 \pi M^2 
  + \frac{24}{\pi } \mu M^2 (\log (\mu ) - 1)
  - \frac{6 }{\pi } \mu^2 M^2 
  + O(\mu^4).
  \label{Eq:SBexp}
\end{eqnarray}

Quantum corrections to the entropy of various black holes were calculated 
using the Cardy formula \cite{Cardy:1986ie,Bloete:1986qm}.  These 
corrections allow us to fix the exact value of $\mu$.  Comparing our 
expression with Refs. \cite{Kaul:2000kf,Carlip:2000nv,Das:2000bx}, we find 
that the Bose case easily lends itself for obtaining a logarithmic 
correction to the black hole entropy.  Setting
\begin{eqnarray}
  \mu = \frac{\pi}{16 M^2},
\end{eqnarray}
the black hole entropy derived from the dual Bose gas becomes 
\begin{eqnarray}
S = 4\,\pi \,M^2 
 - \frac{3}{2} \, \log (4\, \pi M^2)
 - \frac{3}{2} \left( 1+ \log \left( \frac{4}{\pi^2}\right) \right) 
 - \frac{3\,\pi } {128\,M^2} 
 + O(1/M^3).
\end{eqnarray}
We find it remarkable that our simplest toy model, namely a dual one 
dimensional Bose gas, is able to reproduce not only the correct 
semi classical black hole energy and entropy but also their highly 
non-trivial and minute logarithmic quantum corrections. 
It is possible to obtain a value of $\mu$ for the
Fermi gas as well which would be consistent with logarithmic
correction to the entropy.

\section{Conclusions}

We proposed a dual thermodynamics for an isolated
Schwarzschild black hole based on our conjectured energy-entropy duality.
Our new formulation has the major advantage of obeying all laws of 
thermodynamics. This means that ordinary quantum statistics
can represent the otherwise mysterious holography encoded
in a black hole. While it is conceivable that that there are
many dual quantum systems which would be equally suitable for this
purpose, we present toy models based on one dimensional Fermi
or Bose gases. We show that these exactly
reproduce the holographic thermodynamics of black holes.
In particular, since energy and entropy play dual roles, the holographic 
entropy cut off in real space corresponds to the energy
cut off in dual space from the Fermi-Dirac or Bose-Einstein distribution
and vice versa, as it was conjectured in Ref. \cite{Balazs:2006kc}.

The energy-entropy duality transforms a gravitating, strongly interacting, 
low temperature system into a weakly interacting, high temperature dual 
quantum system. This weakly interacting system obeys the laws of ordinary 
quantum mechanics. The duality transformation allows the interpretation of 
results obtained in dual space where calculations are straightforward. We 
have demonstrated this by presenting quantum corrections to the black hole 
mechanics in the framework of our toy model. We find it remarkable that our 
simple model not only reproduces black hole mechanics, but it is also 
consistent with  previously obtained logarithmic corrections to the entropy.

The most conservative interpretation of our result $p dV = 0$ would be that 
the pressure vanishes, $p = 0$. This is consistent with the integral form $E 
= 2 TS$, as can be easily checked from the explicit expressions of energy, 
entropy, and temperature. Our quantum holographic model reproduces all these 
quantities as well as their relationship. However, in extensive 
thermodynamics $E = TS - pV $ should hold if the pressure is non-zero. A 
literal interpretation of our dual model suggests a non-zero negative 
pressure $p = -TS/V$, yielding $E = TS-pV = 2 TS$, consistently with black 
hole mechanics. This pressure term, however, does not enter into the 
(differential) first law of black hole mechanics due to $dL=0$ in our 
underlying quantum model. Thus, taken at face value, this picture predicts 
that a black hole, which classically has zero temperature and pressure, has 
both a non-zero temperature and pressure associated with it. Although this 
pressure does not manifest itself in the differential black hole mechanics 
due to a constraint ($S$, and $V$ are not independent variables), it appears 
in the integral of the first law. Just as the temperature associated with 
black hole mechanics, this pressure is entirely of quantum nature and has an 
equation of state $w = -\frac{1}{2}$, the hallmark of dark energy.

The similarity of Friedmann universes to 
black holes raises the possibility of applying similar ideas toward the 
development of quantum cosmology. In particular, it is intriguing that 
negative pressure is entirely natural in 
this context. Further research into this area could shed light on the 
dark energy component of the universe, and thereby touch base
with observations.  In \cite{Balazs:2006kc}, based 
on simple thermodynamical considerations, we found that the present value of 
the cosmological constant is natural in the holographic context. 
We conjecture that, possibly when ``bulk viscosity'' effects corresponding to
holographic entropy production are taken into account, this negative
quantum gravitational pressure might account for the apparent
acceleration of the universe \cite{Balazs:2006xx}.

The energy-entropy duality as defined in Eq.~\ref{eq:eduality} establishes
a transformation among partition functions as well.
For example, the transformation of the grand canonical partition function
should be
\begin{equation}
  q \equiv \frac{pV}{T} \to -q' T'.
\end{equation}
Exploring the partition functions should shed more light on
the density of states for a black hole. Further investigation
of this issue is left for future research.

It is clear that our considerations can be generalized for other types of 
black holes; e.g., angular momentum and charge add further terms into the 
dual thermodynamics. At the moment general relativity is taken into
account only in a fairly implicit way, through the phenomenology
of black hole mechanics. However, derivation of Einsteins's
equation exists directly from on holographic thermodynamics 
\cite{Jacobson:1995ab}. This
raises the possibility of covariant formulation of these ideas,
possibly based on the work in Ref. \cite{Bousso:1999xy}.
We speculate that this could lead to an
energy-entropy dual of Einstein's equation.

%
It is also clear that a range of dual models could reproduce 
the same large $M$ asymptotic thermodynamics, and possibly
different models could produce 
different quantum effects. The present range of models can be characterized 
by $\mu$ and the choice of Fermi or Bose statistics, however, it is likely 
that other weakly interacting models could be producing similar results. 
Some of them, such as Ising type spin network models, could  have
more direct connection with previous work in this area, although
one could hardly miss the striking similarity of one dimensional
quantum gases with string excitations. Indeed, string theory
appears to have a thermodynamic duality $T \to 1/T$ according
to Ref. \cite{Dienes:2003sq,Dienes:2003dv} and our dual description
is similar to the well known AdS/CFT duality \cite{Maldacena:1997re}. 
We will explore the range possible models and
connections with existing formulations in the future.

Ultimately, it would be desirable to extend these ideas 
to the interaction of a black hole 
with its surrounding. It is might be possible to relate the entropy current 
of the dual system to the energy current of the surrounding of the original 
system, and vice versa. Such calculations could shed more light on
the information paradox associated with the Hawking radiation, and
are left for future research. 

\begin{acknowledgments}

We thank Nick Kaiser, Arjun Menon and Robert Wald for stimulating 
discussions. IS was supported by NASA through AISR NAG5-11996, and ATP NASA 
NAG5-12101 as well as by NSF grants AST02-06243, AST-0434413 and ITR 
1120201-128440. Research at the HEP Division of ANL is supported in part by 
the US DOE, Division of HEP, Contract W-31-109-ENG-38. CB also thanks the 
Aspen Center for Physics for its hospitality and financial support.

\end{acknowledgments}

\bibliographystyle{JHEP}
\bibliography{ms}

\providecommand{\href}[2]{#2}\begingroup\raggedright\begin{thebibliography}{10}

\bibitem{Weinberg:1988cp}
S.~Weinberg, {\it The cosmological constant problem},  {\em Rev. Mod. Phys.}
  {\bf 61} (1989) 1--23.

\bibitem{Riess:1998cb}
{\bf Supernova Search Team} Collaboration, A.~G. Riess {\em et~al.}, {\it
  Observational evidence from supernovae for an accelerating universe and a
  cosmological constant},  {\em Astron. J.} {\bf 116} (1998) 1009--1038,
  [\href{http://xxx.lanl.gov/abs/astro-ph/9805201}{{\tt astro-ph/9805201}}].

\bibitem{Balazs:2006kc}
C.~Balazs and I.~Szapudi, {\it Naturalness of the vacuum energy in holographic
  theories},  \href{http://xxx.lanl.gov/abs/hep-th/0603133}{{\tt
  hep-th/0603133}}.

\bibitem{Bekenstein:1973ur}
J.~D. Bekenstein, {\it Black holes and entropy},  {\em Phys. Rev.} {\bf D7}
  (1973) 2333--2346.

\bibitem{Bekenstein:1974ax}
J.~D. Bekenstein, {\it Generalized second law of thermodynamics in black hole
  physics},  {\em Phys. Rev.} {\bf D9} (1974) 3292--3300.

\bibitem{Bardeen:1973gs}
J.~M. Bardeen, B.~Carter, and S.~W. Hawking, {\it The four laws of black hole
  mechanics},  {\em Commun. Math. Phys.} {\bf 31} (1973) 161--170.

\bibitem{Hawking:1974rv}
S.~W. Hawking, {\it Black hole explosions},  {\em Nature} {\bf 248} (1974)
  30--31.

\bibitem{Hawking:1974sw}
S.~W. Hawking, {\it Particle creation by black holes},  {\em Commun. Math.
  Phys.} {\bf 43} (1975) 199--220.

\bibitem{Susskind:1993if}
L.~Susskind, L.~Thorlacius, and J.~Uglum, {\it The stretched horizon and black
  hole complementarity},  {\em Phys. Rev.} {\bf D48} (1993) 3743--3761,
  [\href{http://xxx.lanl.gov/abs/hep-th/9306069}{{\tt hep-th/9306069}}].

\bibitem{Bekenstein:1972tm}
J.~D. Bekenstein, {\it Black holes and the second law},  {\em Nuovo Cim. Lett.}
  {\bf 4} (1972) 737--740.

\bibitem{'tHooft:1993gx}
G.~'t~Hooft, {\it Dimensional reduction in quantum gravity},
  \href{http://xxx.lanl.gov/abs/gr-qc/9310026}{{\tt gr-qc/9310026}}.

\bibitem{Susskind:1994vu}
L.~Susskind, {\it The world as a hologram},  {\em J. Math. Phys.} {\bf 36}
  (1995) 6377--6396, [\href{http://xxx.lanl.gov/abs/hep-th/9409089}{{\tt
  hep-th/9409089}}].

\bibitem{Bousso:1999xy}
R.~Bousso, {\it A covariant entropy conjecture},  {\em JHEP} {\bf 07} (1999)
  004, [\href{http://xxx.lanl.gov/abs/hep-th/9905177}{{\tt hep-th/9905177}}].

\bibitem{Bousso:2002ju}
R.~Bousso, {\it The holographic principle},  {\em Rev. Mod. Phys.} {\bf 74}
  (2002) 825--874, [\href{http://xxx.lanl.gov/abs/hep-th/0203101}{{\tt
  hep-th/0203101}}].

\bibitem{Wald:1999vt}
R.~M. Wald, {\it The thermodynamics of black holes},  {\em Living Rev. Rel.}
  {\bf 4} (2001) 6, [\href{http://xxx.lanl.gov/abs/gr-qc/9912119}{{\tt
  gr-qc/9912119}}].

\bibitem{Banks:1983by}
T.~Banks, L.~Susskind, and M.~E. Peskin, {\it Difficulties for the evolution of
  pure states into mixed states},  {\em Nucl. Phys.} {\bf B244} (1984) 125.

\bibitem{Wald:1995yp}
R.~M. Wald, {\it Quantum field theory in curved space-time and black hole
  thermodynamics}, . Chicago, USA: Univ. Pr. (1994) 205 p.

\bibitem{Hartle:1996rp}
J.~B. Hartle, {\it Generalized quantum theory in evaporating black hole
  spacetimes},  \href{http://xxx.lanl.gov/abs/gr-qc/9705022}{{\tt
  gr-qc/9705022}}.

\bibitem{Strominger:1996sh}
A.~Strominger and C.~Vafa, {\it Microscopic origin of the bekenstein-hawking
  entropy},  {\em Phys. Lett.} {\bf B379} (1996) 99--104,
  [\href{http://xxx.lanl.gov/abs/hep-th/9601029}{{\tt hep-th/9601029}}].

\bibitem{Peet:1997es}
A.~W. Peet, {\it The bekenstein formula and string theory (n-brane theory)},
  {\em Class. Quant. Grav.} {\bf 15} (1998) 3291--3338,
  [\href{http://xxx.lanl.gov/abs/hep-th/9712253}{{\tt hep-th/9712253}}].

\bibitem{Aharony:1999ti}
O.~Aharony, S.~S. Gubser, J.~M. Maldacena, H.~Ooguri, and Y.~Oz, {\it Large n
  field theories, string theory and gravity},  {\em Phys. Rept.} {\bf 323}
  (2000) 183--386, [\href{http://xxx.lanl.gov/abs/hep-th/9905111}{{\tt
  hep-th/9905111}}].

\bibitem{Carlip:2000nv}
S.~Carlip, {\it Logarithmic corrections to black hole entropy from the cardy
  formula},  {\em Class. Quant. Grav.} {\bf 17} (2000) 4175--4186,
  [\href{http://xxx.lanl.gov/abs/gr-qc/0005017}{{\tt gr-qc/0005017}}].

\bibitem{Carlip:2006fm}
S.~Carlip, {\it Horizons, constraints, and black hole entropy},
  \href{http://xxx.lanl.gov/abs/gr-qc/0601041}{{\tt gr-qc/0601041}}.

\bibitem{Mathur:2005zp}
S.~D. Mathur, {\it The fuzzball proposal for black holes: An elementary
  review},  {\em Fortsch. Phys.} {\bf 53} (2005) 793--827,
  [\href{http://xxx.lanl.gov/abs/hep-th/0502050}{{\tt hep-th/0502050}}].

\bibitem{Ashtekar:1997yu}
A.~Ashtekar, J.~Baez, A.~Corichi, and K.~Krasnov, {\it Quantum geometry and
  black hole entropy},  {\em Phys. Rev. Lett.} {\bf 80} (1998) 904--907,
  [\href{http://xxx.lanl.gov/abs/gr-qc/9710007}{{\tt gr-qc/9710007}}].

\bibitem{Livine:2005mw}
E.~R. Livine and D.~R. Terno, {\it Quantum black holes: Entropy and
  entanglement on the horizon},  {\em Nucl. Phys.} {\bf B741} (2006) 131--161,
  [\href{http://xxx.lanl.gov/abs/gr-qc/0508085}{{\tt gr-qc/0508085}}].

\bibitem{Frolov:1997up}
V.~P. Frolov and D.~V. Fursaev, {\it Mechanism of generation of black hole
  entropy in sakharov's induced gravity},  {\em Phys. Rev.} {\bf D56} (1997)
  2212--2225, [\href{http://xxx.lanl.gov/abs/hep-th/9703178}{{\tt
  hep-th/9703178}}].

\bibitem{Hawking:1998jf}
S.~W. Hawking and C.~J. Hunter, {\it Gravitational entropy and global
  structure},  {\em Phys. Rev.} {\bf D59} (1999) 044025,
  [\href{http://xxx.lanl.gov/abs/hep-th/9808085}{{\tt hep-th/9808085}}].

\bibitem{Wald:1997qp}
R.~M. Wald, {\it The *nernst theorem* and black hole thermodynamics},  {\em
  Phys. Rev.} {\bf D56} (1997) 6467--6474,
  [\href{http://xxx.lanl.gov/abs/gr-qc/9704008}{{\tt gr-qc/9704008}}].

\bibitem{Cardy:1986ie}
J.~L. Cardy, {\it Operator content of two-dimensional conformally invariant
  theories},  {\em Nucl. Phys.} {\bf B270} (1986) 186--204.

\bibitem{Bloete:1986qm}
H.~W.~J. Bloete, J.~L. Cardy, and M.~P. Nightingale, {\it Conformal invariance,
  the central charge, and universal finite size amplitudes at criticality},
  {\em Phys. Rev. Lett.} {\bf 56} (1986) 742--745.

\bibitem{Kaul:2000kf}
R.~K. Kaul and P.~Majumdar, {\it Logarithmic correction to the
  bekenstein-hawking entropy},  {\em Phys. Rev. Lett.} {\bf 84} (2000)
  5255--5257, [\href{http://xxx.lanl.gov/abs/gr-qc/0002040}{{\tt
  gr-qc/0002040}}].

\bibitem{Das:2000bx}
S.~Das, R.~K. Kaul, and P.~Majumdar, {\it A new holographic entropy bound from
  quantum geometry},  {\em Phys. Rev.} {\bf D63} (2001) 044019,
  [\href{http://xxx.lanl.gov/abs/hep-th/0006211}{{\tt hep-th/0006211}}].

\bibitem{Balazs:2006xx}
C. Balazs, N. Kaiser and I. Szapudi in preparation.

\bibitem{Jacobson:1995ab}
T.~Jacobson, {\it Thermodynamics of space-time: The einstein equation of
  state},  {\em Phys. Rev. Lett.} {\bf 75} (1995) 1260--1263,
  [\href{http://xxx.lanl.gov/abs/gr-qc/9504004}{{\tt gr-qc/9504004}}].

\bibitem{Dienes:2003sq}
K.~R. Dienes and M.~Lennek, {\it Thermal duality confronts entropy: A new
  approach to string thermodynamics?},
  \href{http://xxx.lanl.gov/abs/hep-th/0312173}{{\tt hep-th/0312173}}.

\bibitem{Dienes:2003dv}
K.~R. Dienes and M.~Lennek, {\it Adventures in thermal duality. ii: Towards a
  duality- covariant string thermodynamics},  {\em Phys. Rev.} {\bf D70} (2004)
  126006, [\href{http://xxx.lanl.gov/abs/hep-th/0312217}{{\tt
  hep-th/0312217}}].

\bibitem{Maldacena:1997re}
J.~M. Maldacena, {\it The large n limit of superconformal field theories and
  supergravity},  {\em Adv. Theor. Math. Phys.} {\bf 2} (1998) 231--252,
  [\href{http://xxx.lanl.gov/abs/hep-th/9711200}{{\tt hep-th/9711200}}].

\end{thebibliography}\endgroup


\end{document}